\documentclass[10pt,letterpaper]{article}

\usepackage{ccn}
\usepackage{pslatex}
\usepackage{apacite}
\usepackage{graphicx}

\title{Cortical Mirror-System Activation During Real-Life Game Playing: \newline An Intracranial Electroencephalography (EEG) Study}
 
\author{{\large \bf Markus Kern (kern@uniklinik-freiburg.de)} \\
BrainLinks-BrainTools Cluster of Excellence, University of Freiburg, Germany
 \AND {\large \bf Johanna Ruescher (hanna-we@gmx.de)} \\
  Faculty of Biology – University of Freiburg\\
  79104, Freiburg, Germany
 \AND {\large \bf Andreas Schulze-Bonhage (schulze-bonhage@uniklinik-freiburg.de)} \\
  Epilepsy Center, Department of Neurosurgery, Medical Center — University of Freiburg,\\
  79106, Freiburg, Germany
 \AND {\large \bf Tonio Ball (tonio.ball@uniklinik-freiburg.de)} \\
Medical AI Lab, University Medical Center Freiburg \\ 
79106, Freiburg, Germany}

\begin{document}

\maketitle

\section{Abstract}
{
\bf
Analogous to the mirror neuron system repeatedly described in monkeys as a possible substrate for imitation learning and/or action understanding, a neuronal execution/observation matching system (OEMS) is assumed in humans, but little is known to what extent this system is activated in non-experimental, real-life conditions. In the present case study, we investigated brain activity of this system during natural, non-experimental motor behavior as it occurred during playing of the board game "Malefiz". We compared spectral modulations of the high-gamma band related to ipsilateral reaching movement execution and observation of the same kind of movement using electrocorticography (ECoG) in one participant. Spatially coincident activity during both conditions execution and observation was recorded at electrode contacts over the premotor/primary motor cortex. The topography and amplitude of the high-gamma modulations related to both, movement observation and execution were clearly spatially correlated over several fronto-parietal brain areas. Thus, our findings indicate that a network of cortical areas contributes to the human OEMS, beyond primary/premotor cortex including Broca’s area and the temporo-parieto-occipital junction area, in real-life conditions.\newline 

Keywords: mirror system; ECoG; non-experimental; real life
}

\section{Introduction}

Mirror neurons that respond both to (visually or auditory) movement observation and execution, have first been described in monkeys in area F5 of the ventral premotor cortex \cite{pellegrino_understanding_1992,rizzolatti_localization_1996}, in the inferior parietal cortex (IPC) \cite{fogassi_parietal_2005} and the intraparietal sulcus (IPS) \cite{keysers_mirror_2009}. The analogous observation-execution matching system (OEMS) in humans, which is suggested to play a crucial role in imitation learning \cite{iacoboni_beyond_2006} and/or action understanding \cite{rizzolatti_premotor_1996}, is assumed to be more widely distributed over the human cortex than in monkeys \cite{buccino_action_2001,keysers_social_2010}. Indications of Human brain activity associated with the OEMS were observed in several cortical areas including the superior temporal sulcus (STS) \cite{rizzolatti_localization_1996}, Broca´s area (BR) \cite{gazzola_observation_2009,iacoboni_cortical_1999,rizzolatti_localization_1996}, the IPC \cite{gazzola_observation_2009,newman-norlund_mirror_2007}, the dorsal premotor cortex  \cite{gazzola_observation_2009}, the ventral premotor cortex (vPM) \cite{buccino_action_2001,dinstein_brain_2007,filimon_observed_2015},  the supplementary motor area (SMA) \cite{gazzola_observation_2009,hamzei_human_2003,iacoboni_cortical_1999}, the superior parietal cortex (SPC) \cite{gazzola_observation_2009,iacoboni_cortical_1999}, as well as the primary and secondary somatosensory cortex (S1/S2) \cite{gazzola_observation_2009}.

However, while these experimental studies clearly show the existence of an OEMS in the human brain, it is not clear whether the same kind of overlapping brain activities are also present during non-experimental, real-life conditions. Since striking differences in the firing behavior of cortical cells of the monkey brain were observed during natural movements compared to the instructed ones during experimental conditions \cite{harrison_motor_2014}, the investigation of human brain activity in real-life conditions is of interest.

Thus, in the present study we collected ECOG data during natural movement execution and observation, i.e., while the patient was engaged in the board game “Malefiz” with her life partner. This particular situation was chosen for due to the following reasons: i) the high degree of uniformity of arm movements in the observation and execution condition, ii) in the playing situation, the visitor was sitting close enough to the patients’ bed to be filmed by the digital video recorded for medical reasons, capturing only the direct vicinity of the hospital bed and iii) the patients’ attention during observation trials was specifically concentrated on the movements/actions of her game partner. 
\begin{figure}[ht]
\begin{center}
\includegraphics[width=0.48\textwidth]{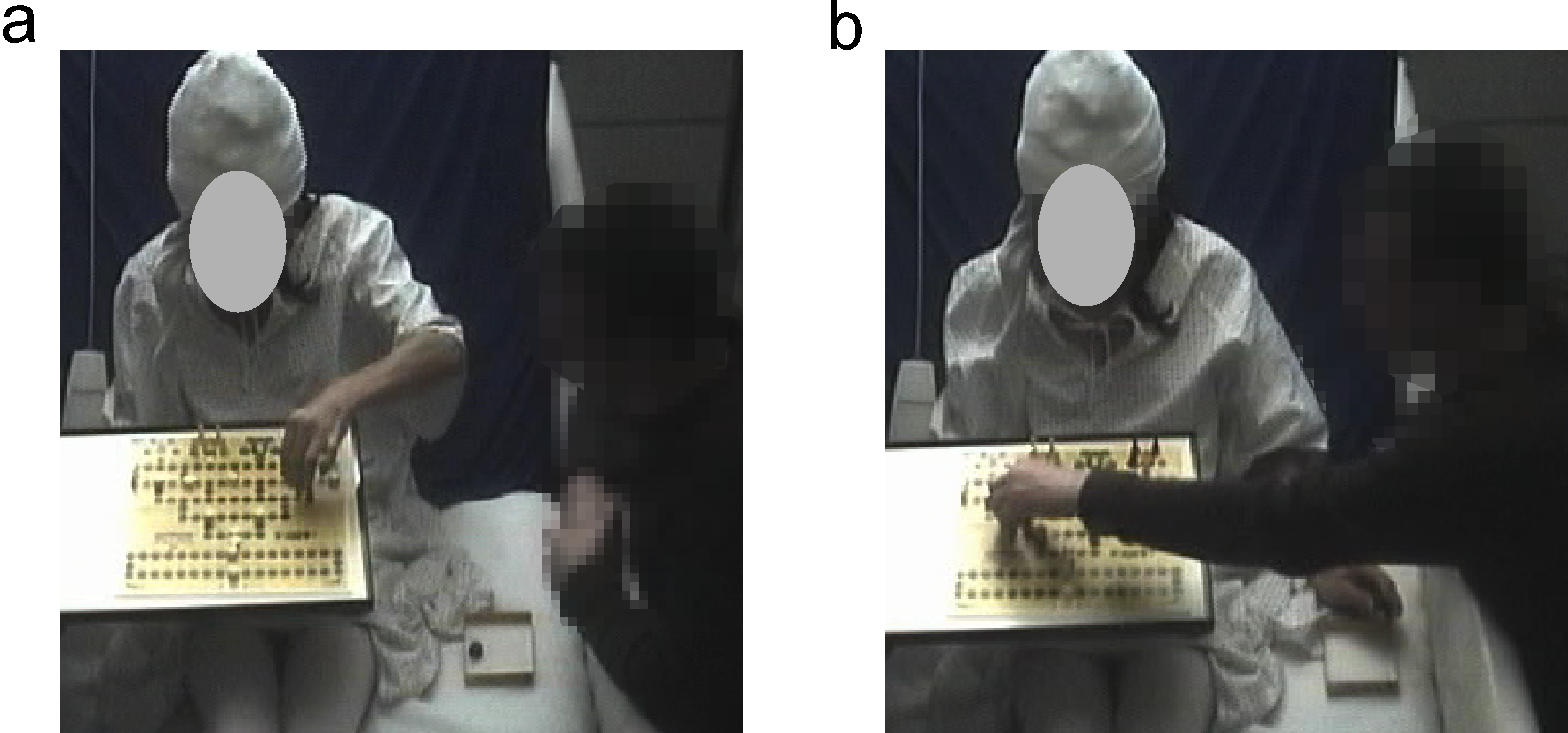}
\end{center}
\caption{Game situation analyzed in this study. a) Movement execution condition b) Movement observation condition (both video frames anonymized).} 
\label{figgame}
\end{figure}

\section{Methods}

Data from a female, left handed 41 year-old patient were used, who was suffering from pharmacological resistant epilepsy and had a subdural 8*8 platinum electrode grid implanted over the left fronto-parietal cortex for pre-surgical evaluation. The patient agreed by informed consent to scientific usage of the recorded data and the study was approved by the local ethics committee. ECoG was recorded with a sampling rate of 1024 Hz. Monopolar electrical stimulation was conducted by physicians of the University Hospital Freiburg using an INOMED NS 60 stimulator (INOMED, Germany) (Fig. 2a). Details about the stimulation procedure are described in \cite{ruescher_somatotopic_2013}. Anatomical assignment of electrodes to functional brain areas have been performed using the hierarchic anatomical assignment approach described in \cite{pistohl_decoding_2012}. 

We analyzed the ECoG data recorded while the patient played the game “Malefiz” with her life partner. In this game, both partners performed similar reaching movements to move their tokens on the game board. Due to handedness (left) and hemisphere of implantation (left), we analyzed ipsilateral reaching movements of the patient in the execution condition, consisting of 120 trials (Fig.1a). In the observation condition, we analyzed 87 trials of the left (14 trials) and right (73 trials) arm of the patient’s visitor (Fig.1b). Both persons were monitored by digital video recorded for medical reasons with a 25-Hz sampling rate and 640x480 pixels resolution. Trials were pre-selected by visual inspection of video recordings and confirmed by the analysis of the EMG recordings of the right and left deltoid muscles of the patient.

Event-related spectral power was calculated using a multi-taper spectral analysis method \cite{percival_wavelet_2006} with time windows of 200 ms, 20-ms time steps and 5 Slepian tapers. Since event-related HG power is proposed to be a suitable marker of local brain activity \cite{crone_functional_1998}, we analyzed modulations of HG band power relative to the power in baseline trials (resting condition, no movement execution or observation of movement, 21 trials).

The significance of HG power modulations was calculated using a paired two-sided Wilcoxon’s sign test. P-values were corrected by using the false discovery rate (FDR) approach \cite{benjamini_controlling_2001} with a q-level of $10^{-9}$. For a comparison of the overall fronto-parietal brain activity captured by grid electrodes during both conditions, we calculated Spearman’s rank-based correlations coefficient for HG power related between the execution and observation condition over the 64 grid electrode contacts.

\section{Results}

Both the execution and observation condition showed widespread high gamma (HG) band modulations in frontal and parietal regions (Fig. 2b,c). For the execution condition (Fig. 2b), significant increases of HG power were found over S1, PM, and IPC, while significant HG power decreases were found at two electrodes above the central sulcus and in the ventral S1. In the observation condition (Fig. 2c), significant HG power increases were found in PM, temporo-parieto-occipital junction area (TPO), dorsal S1 and BR.

On the single electrode level, significant responses overlapped in dorsal premotor cortex in a region showing hand motor responses during electrical stimulation mapping (Fig. 2a). Comparison of HG response topographies recorded over the whole electrode grid revealed significant positive correlations with r=0.71 and $p<10^{-6}$ (Fig. 2d).

\section{Discussion}

In the present case study, we investigated the human OEMS by analyzing HG modulations in ECoG recordings of a female epilepsy patient while she was playing a board game with her partner. We found a significant increase of HG power at electrodes over PM during ipsilateral reaching movements as well as during the observation of unilateral reaching movements of the partner. This coincidental activity in PM is in line with results of recent fMRI studies investigating the OEMS \cite{gazzola_observation_2009}.

Overlapping activity was also recorded over Broca’s Area, the temporo-parietö-occipital (TPO) junction area and dorsal S1, but did not reach significance in the execution condition. This might partially be ascribed to the laterality of movement execution, since a movement-related increase in HG power is rather weak in the ipsilateral hemisphere compared to the contralateral side. However, BR and S1 have also been associated to the OEMS in previous studies \cite{gazzola_observation_2009,iacoboni_cortical_1999,rizzolatti_localization_1996}.

\begin{figure*}[h!t]
\centerline{\includegraphics[width=\textwidth]{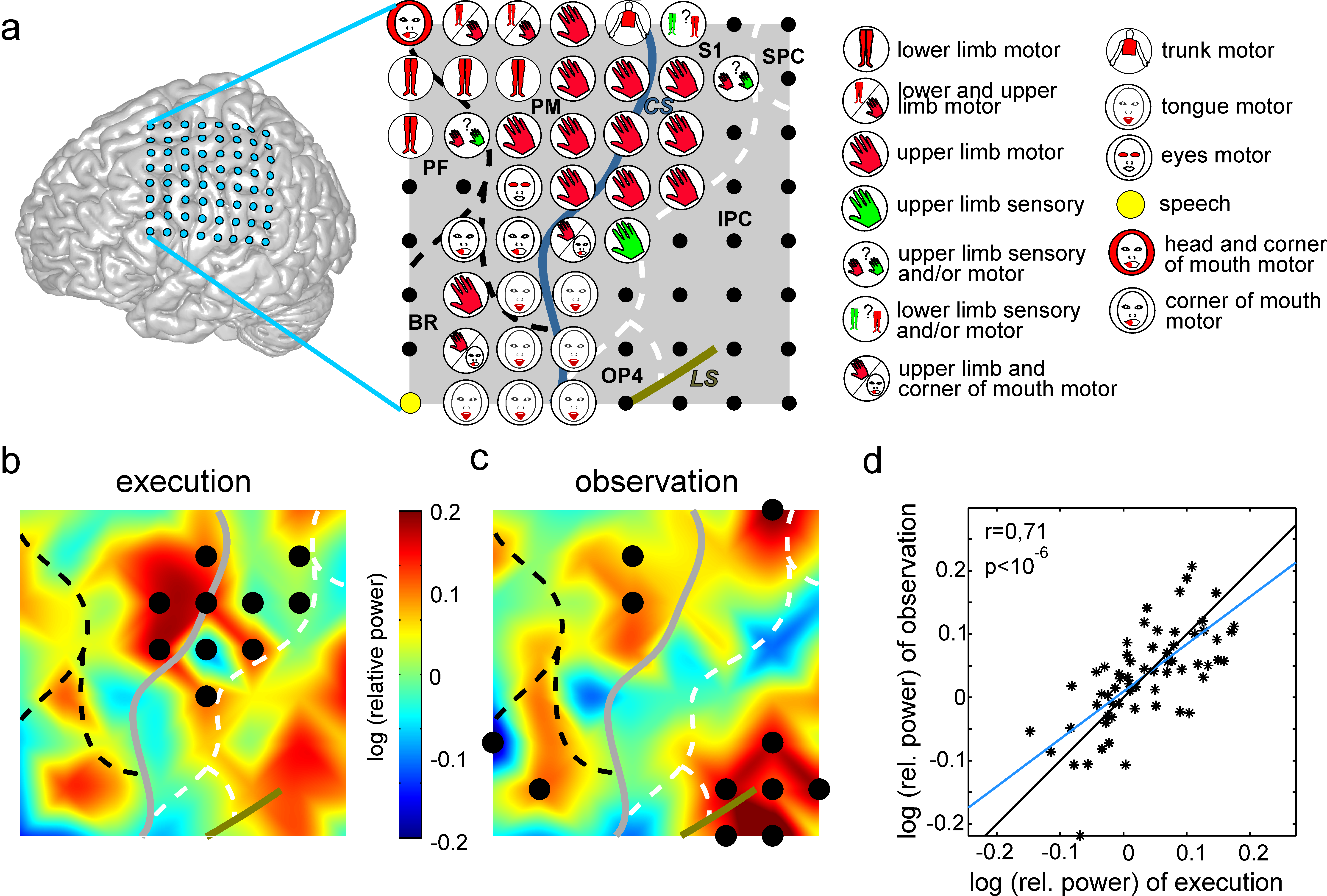}}
\caption{Electrical stimulation mapping (ESM) and high gamma (HG) mapping results. a) Left panel: position of 8*8 subdurally implanted electrode grid of the patient visualized on a standard brain; right panel: anatomical assignment of the electrode contacts and responses during ESM. Prefrontal cortex: PF, Broca’s area: BR, premotor cortex: PM, primary somatosensory cortex: S1, inferior parietal cortex: IPC, superior parietal cortex: SPC. Solid blue line: central sulcus (CS), solid green line: lateral sulcus (LS), dashed black lines: probabilistically defined borders within frontal cortex, dashed white line: probabilistically defined border between the anterior and posterior parietal cortex. Electrode contacts marked by black dots were not tested or did not elicit responses during ESM. b) Topography of reaching-execution-related relative power changes in the HG band. Electrodes where significant (sign test, $p<10^{-9}$, FRD corrected) power changes were recorded, are marked as black dots. c) Topography of reaching-observation-related relative power changes (color bar) in high-gamma band. Conventions as in b) d) The logarithmic relative HG band power values (black stars) of reaching observation and execution in the time window 0-500ms displayed in (b,c) are plotted against each other. Blue line: regression line. Correlation coefficients and p-values after Spearman´s correlation are displayed in the plot.} 
\label{figmapping}
\end{figure*}

However, BR and S1 have also been associated to the OEMS in previous studies \cite{gazzola_observation_2009,iacoboni_cortical_1999,rizzolatti_localization_1996}. TPO has been found to be active during action observation \cite{hamzei_human_2003} and might thus play a role in the OEMS, too.

Additionally, we found significant coincident activity in S1, PM, prefrontal cortex and BR in another epilepsy patient not included in this study, related to contralateral finger movement execution (for more details, see P1 in \cite{ball_movement_2008}), compared to the observation of her partner manipulating a Rubik’s cube.

Taken together, these findings support the suggestion, that the coincident activity we recorded over PM, BR, TPO and S1 during real-life conditions may be ascribed to the human OEMS. Interestingly, when comparing HG power at every channel of the fronto-parietal electrode grid, we found strong correlations between HG band modulations related to both conditions. First of all, this finding confirms the idea of multiple brain areas contributing to the human OEMS. Furthermore, the similarity of recorded signals during movement observation and execution might also be of interest for future research in the area of brain-computer interfaces, where it is crucial to reliably differentiate between these two conditions. 

\section{Acknowledgments}

We would like to thank Simon Contzen for his help in preparing this manuscript. This study was (partly) funded by the WISPER Project, Baden-Württemberg Stiftung.

\bibliographystyle{apacite}

\setlength{\bibleftmargin}{.125in}
\setlength{\bibindent}{-\bibleftmargin}

\bibliography{main}

\end{document}